\documentclass[10pt]{article}

\usepackage[utf8]{inputenc} % set input encoding (not needed with XeLaTeX)

\usepackage{geometry} % to change the page dimensions
\usepackage{graphicx} % support the \includegraphics command and options
\usepackage[parfill]{parskip} % Activate to begin paragraphs with an empty line rather than an indent
\usepackage{booktabs} % for much better looking tables
\usepackage{array} % for better arrays (eg matrices) in maths
\usepackage{paralist} % very flexible & customisable lists (eg. enumerate/itemize, etc.)
\usepackage{verbatim} % adds environment for commenting out blocks of text & for better verbatim
\usepackage{subfig} % make it possible to include more than one captioned figure/table in a single float
\usepackage{authblk}
\usepackage{amssymb,amsmath}
\usepackage{color}
\usepackage{fancyhdr} % This should be set AFTER setting up the page geometry
\usepackage{sectsty}
\allsectionsfont{\sffamily\mdseries\upshape} % (See the fntguide.pdf for font help)
\def\n{\chi_1^0}
\def\c{\chi_1^\pm}

\newcommand{\be}{\begin{equation}}
\newcommand{\ee}{\end{equation}}
\newcommand{\bea}{\begin{eqnarray}}
\newcommand{\eea}{\end{eqnarray}}
\newcommand{\bi}{\begin{itemize}}
\newcommand{\ei}{\end{itemize}}

\title{The NMSSM at the Cosmic Frontier for Snowmass 2013}
\author{ Mathew~McCaskey$^1$ and Gabe~Shaughnessy$^2$, \\
$^1$Department of Physics, University of Kansas, Lawrence, KS 66044, USA \\
$^2$Department of Physics, University of Wisconsin, Madison, WI 53706, USA}

\begin{document}

\maketitle
\begin{abstract}
We examine the NMSSM at the cosmic frontier in the post Higgs discovery world.  For DM relic abundance consistent with measurement, we find the neutralino can either be singlino or bino dominated.  Wino and higgsino DM generally yield a lower abundance, but offer opportunities of detection at IceCube.  For both cases, future SI direct detection experiments cover a majority of the model, including nearly all of the bino and higgsino scenarios.
\end{abstract}

The MSSM is arguably one of the most popular models beyond the SM that attempts to solve the hierarchy problem.  However, despite the successes of the MSSM there are still a few shortcomings.  Among them, the MSSM superpotential contains the term
\be
W_{MSSM}\supset \mu H_{u}H_{d}
\ee
\noindent where the scale of $\mu$ is an input parameter that has no a-priori connection to the electroweak scale. The Next-to-Minimal Supersymmetric Standard Model (NMSSM) is a simple extension to the MSSM that aims to alleviate this problem, where the $\mu$ parameter is promoted to a dynamical field.  The singlet terms of the NMSSM superpotential are written as
\be
W_{NMSSM}\supset \lambda_{S}SH_{u}H_{d} + \frac{\kappa}{3}S^{3}
\ee
\noindent where $\lambda_{S}$ and $\kappa$ are dimensionless parameters.  When the singlet obtains a VEV an effective $\mu$ parameter is generated.  The additional scalar of the NMSSM and similar singlet extended SUSY models increases the field content, including offering another dark matter candidate, the singlino, $\tilde S$, which has been extensively studied~\cite{Barger:2005hb}.

\begin{figure}[!t]
\begin{center}
     \includegraphics[angle=0,width=0.32\textwidth]{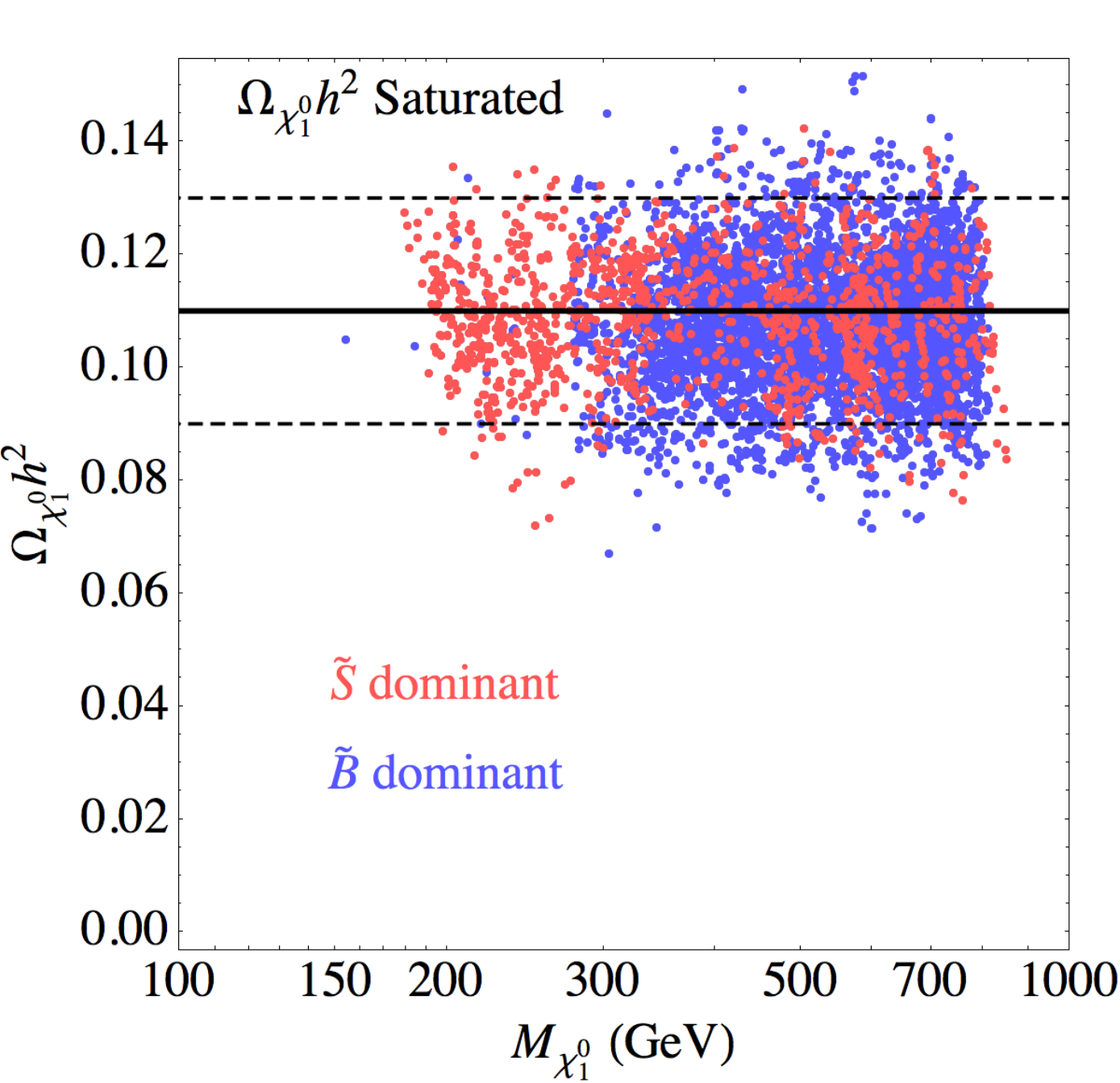}
     \includegraphics[angle=0,width=0.32\textwidth]{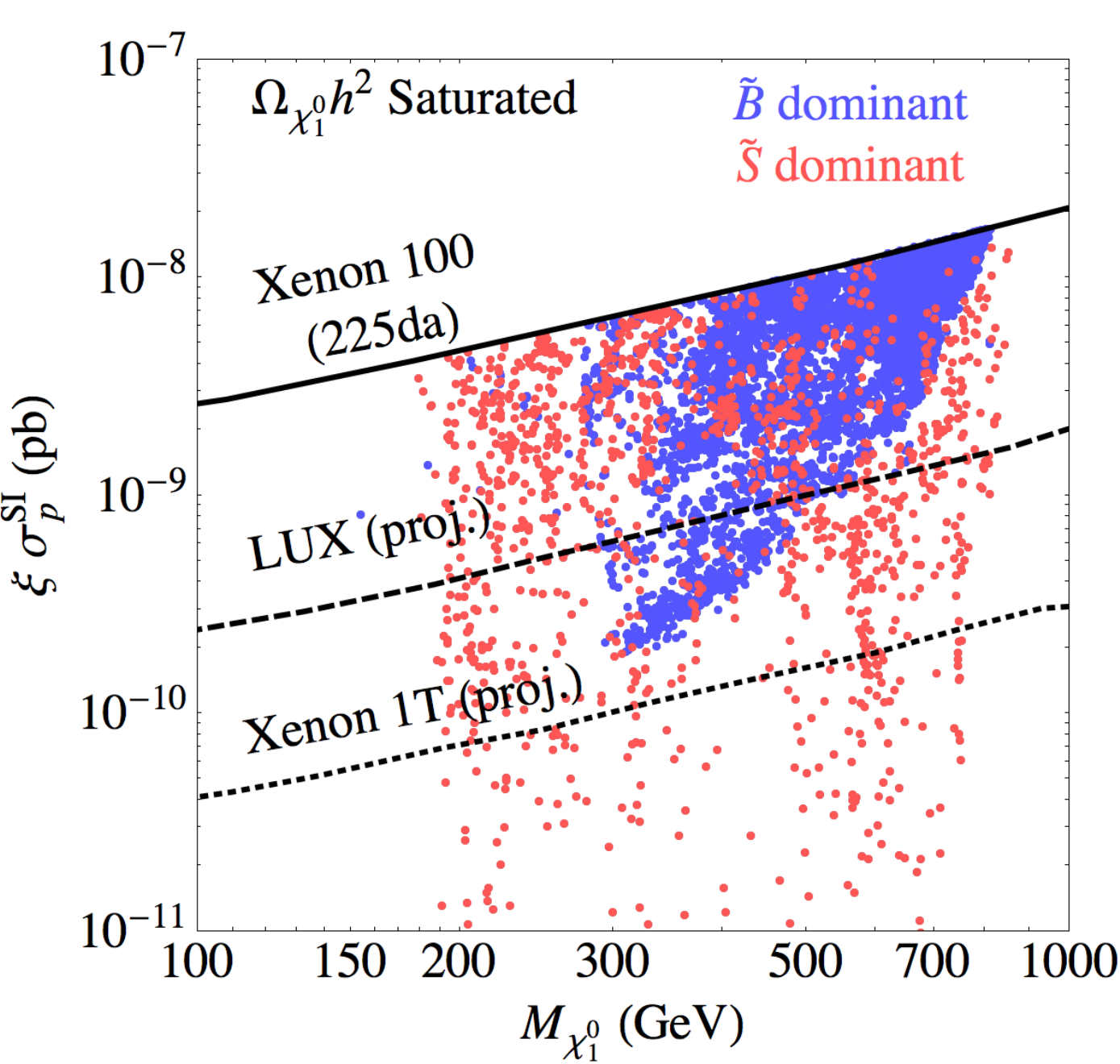}
     \includegraphics[angle=0,width=0.32\textwidth]{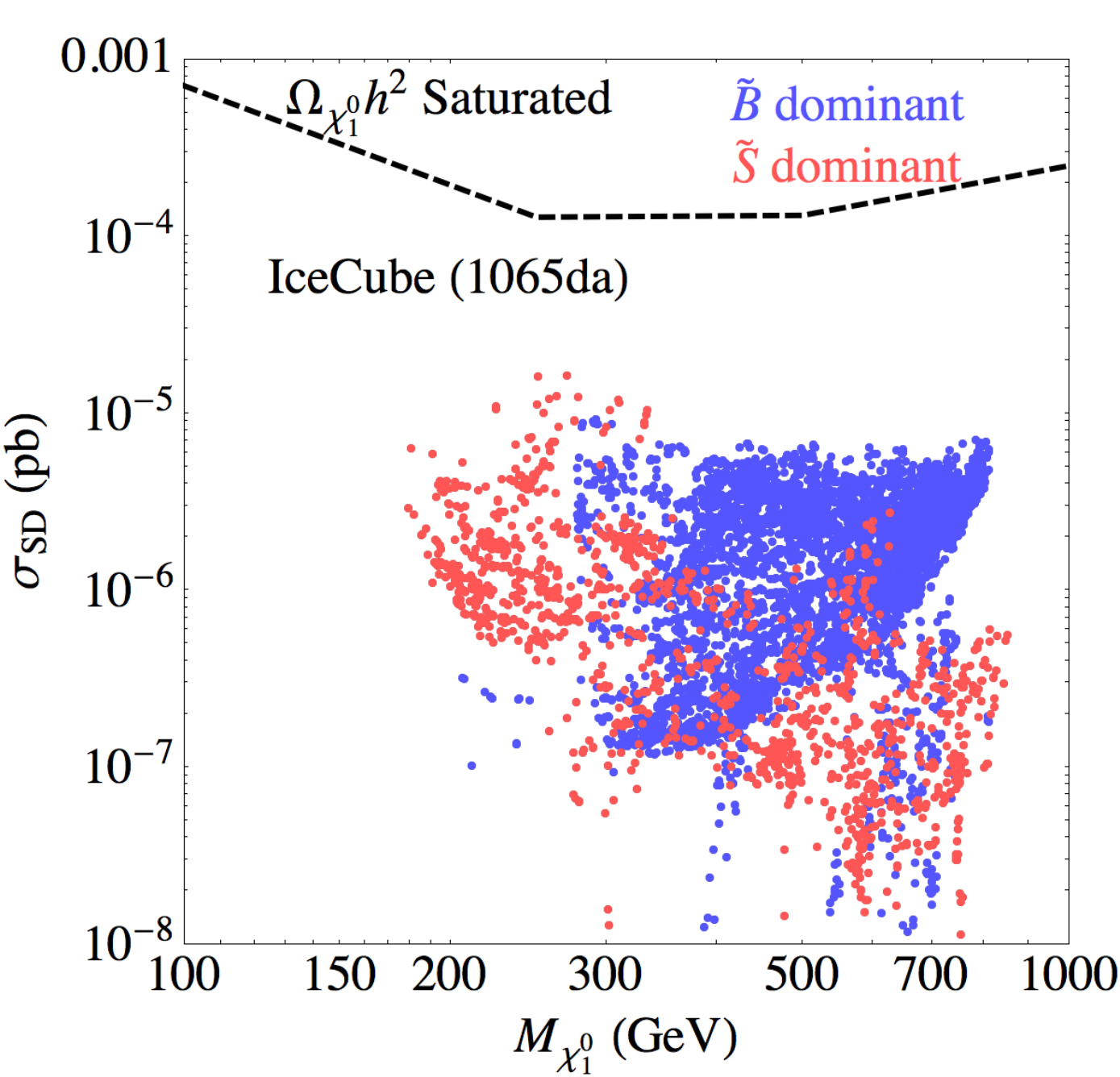}\\
     \includegraphics[angle=0,width=0.32\textwidth]{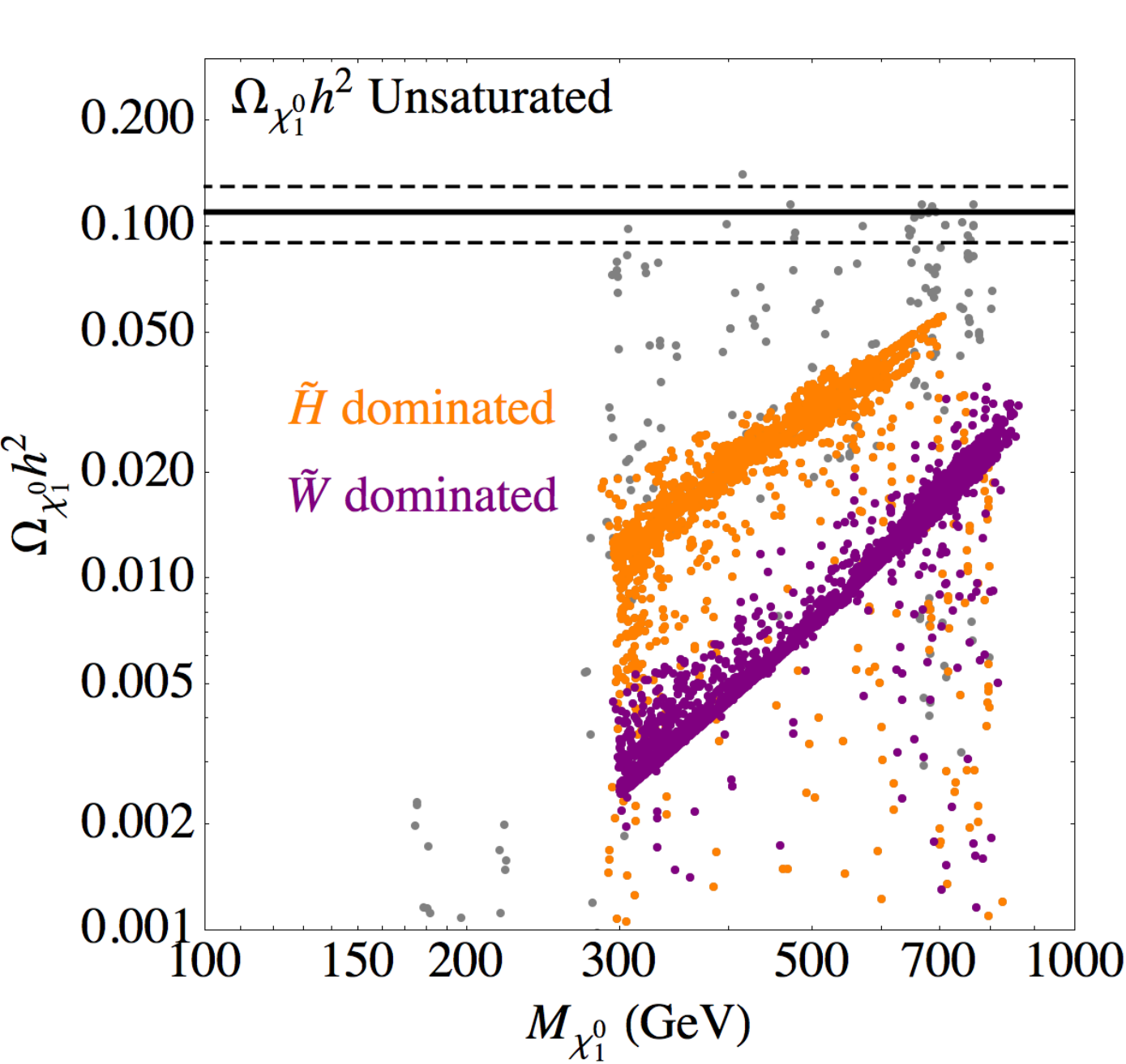}
     \includegraphics[angle=0,width=0.32\textwidth]{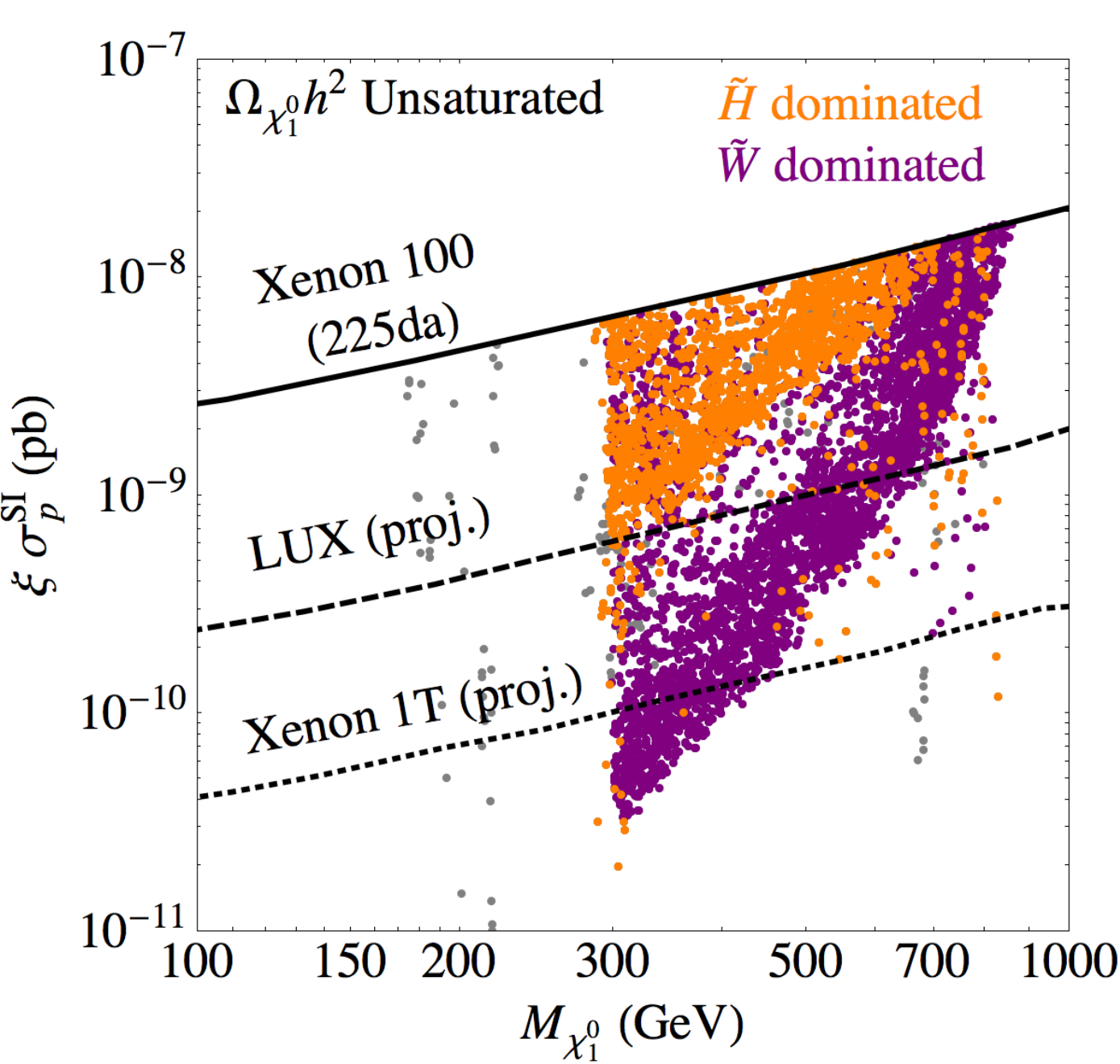}
     \includegraphics[angle=0,width=0.32\textwidth]{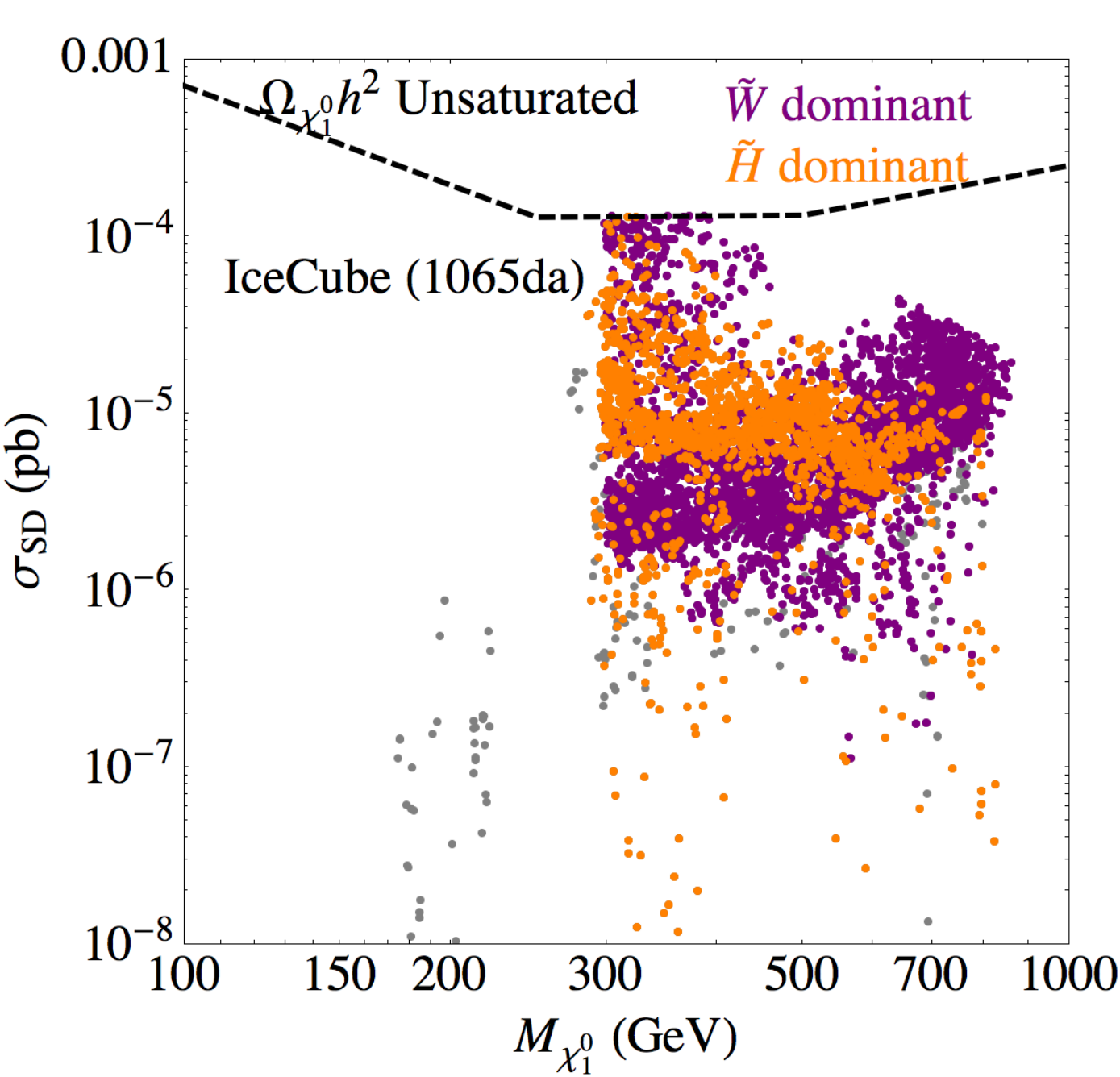}
\caption{Relic abundance (left), scaled SI cross section (middle) and SD cross section (right) for the saturated (top) and unsaturated (bottom) DM relic abundance scenarios.  Limits on $\sigma_{\rm SI}$ from Xenon100~\cite{Aprile:2012nq}, LUX~\cite{Akerib:2012ak}, and Xenon 1 Ton~\cite{Aprile:2012zx}, and $\sigma_{\rm SD}$ from IceCube~\cite{IceCube:2011aj}.  Figures taken from Ref.~\cite{McCaskey:2013xx}.  }
\label{figs}
\end{center}
\end{figure}

To investigate the behavior of the NMSSM with current data, we perform a fit assuming decoupled sfermions to focus on the role of the singlet and singlino.  We explore two cases, where $\n$ can account for the dark matter relic abundance in its entirety, or it is a subcomponent.  In our fit, we require require consistency with SUSY searches~\cite{Ellwanger:2004xm} and present $h$(125) measurements at the LHC~\cite{ATLAS:2013sla}. In the case of a saturated relic abundance, we assume a 10\% theoretical uncertainty on the measured value of $\Omega_{DM}h^2$. The relic abundance for these cases are shown in the left panels of Fig.~\ref{figs}.

%%%%%%%%%%%%%%%%%%%%%%%%%%%%%%%%%%%%%%%%%%%%%%%%
%%%%%%%%%%%%%%%%%%%%%%%%%%%%%%%%%%%%%%%%%%%%%%%%

The LHC $h$(125) measurements indicate a SM-like Higgs boson.  Therefore, if $\tan \beta \gg 1$, the lightest Higgs is typically dominated by $H_u$.  As a consequence, the $h \n\n$ coupling that primarily controls $\sigma_{\rm SI}$ avoids tension with the Xenon100 exclusion if $\n$ is mainly bino or wino.  When $\n$ is a wino or higgsino, the masses $\chi_2^0$ and $\c$ are comparable to $\n$, leading to significant co-annihilation effects.  Moreover, this scenario offers a difficult scenario at the LHC for discovering these states as the decay products will typically be soft and evade the selection cuts.  

In the saturated relic abundance scenario, $\n$ can be either bino or singlino.  Some key observables are illustrated in the top row of Fig.~\ref{figs}.
\bi
\item In the bino scenario, $\sigma v_{\rm tot}$ is often below $3\times 10^{-27}$ cm$^3$ s$^{-1}$, therefore $\n\c$ and $\n\chi_2^0$ co-annihilation is required, with mass splitting of ${\cal O }(20\text{ GeV})$.  The SI scattering cross sections are within reach of future detectors such as Xenon 1 Ton and LUX. 
\item In the singlino scenario, $\sigma_{\rm SI}$ is reduced since usually dominant $h$ exchange is suppressed by the lack of a singlet component.  The singlino case can have a generally smaller $\sigma_{\rm SI}$.
\ei
In either case, the SD cross section is more than a factor of 10 below the current bound placed by COUPP~\cite{Behnke:2012ys} and IceCube~\cite{IceCube:2011aj}.  A light scalar singlet with $M_s < 100$ GeV is possible, which may offer interesting signatures at the LHC such as $h\to ss\to b\bar b + b\bar b (\tau\tau)$.

The unsaturated case is dominated by either higgsino or wino DM.  In both cases, co-annihilation with the associated $\c$ results in a low relic abundance, thus the direct detection sensitivities are scaled by the local density $\rho_{\rm \n} \propto \xi \equiv {\Omega_{\n} h^2/0.11}$.
\bi
\item The higgsino scenario accounts for the upper band of relic density and scaled SI cross section in the correspoding panels in Fig.~\ref{figs}, and is therefore within reach of future detectors such as LUX or Xenon 1 Ton.  
\item A wino dominated $\n$ is more suppressed than the higgsino scenario, and resides in the lower band of the relic abundance and $\sigma_{\rm SI}$ and may fall beyond the reach of Xenon 1 Ton.

\ei
Since the primary exchange for $\sigma_{\rm SD}$ occurs through a $Z$ boson, the cross section is proportional to the higgsino asymmetry.  This asymmetry is generally suppressed for large neutralino masses.  The current experimental sensitivity of COUPP is about an order of magnitude larger than the scaled scattering rate, $\xi \sigma_{\rm SD}$.  However, the sensitivity from IceCube does not critically depend on $\xi$.  Therefore, for masses below 500 GeV, IceCube may soon be sensitive to both higgsino and wino scenarios.

%%%%%%%%%%%%%%%%%%%%%%%%%%%%
\vspace{-5mm}
\section*{Acknowledgements}
\vspace{-4mm}
%%%%%%%%%%%%%%%%%%%%%%%%%%%%
GS is supported by the U. S. Department of Energy under contract DE-FG-02-95ER40896.  MM is supported by U. S. Department of Energy grant DE-FG02-04ER14308 and by National Science Foundation grant PHY-0653250.
%%%%%%%%%%%%%%%%%%%%%%%%%%%%

\end{document}